\documentclass[reprint,
 amsmath,amssymb,
 aps,
]{revtex4-2}

\usepackage{graphicx}
\usepackage{dcolumn}
\usepackage{bm}
\usepackage{hyperref}
\usepackage{siunitx}
\usepackage{spalign}
\usepackage{algorithm}
\usepackage{algpseudocode}

\begin{document}

\preprint{APS/123-QED}

\title{Second Order Meanfield Approximation \\ for calculating Dynamics in $\mathrm{Au}$-Nanoparticle Networks} 

\author{Evan Wonisch}
\author{Jonas Mensing}
\author{Andreas Heuer}
\affiliation{Department for Physical Chemistry, University of Münster, Germany}

\date{\today}

\begin{abstract}
\centering
Exploiting physical processes for fast and energy-efficient computation bears great potential in the advancement of modern hardware components. This paper explores non-linear charge tunneling in nanoparticle networks, controlled by external voltages. The dynamics are described by a master equation, which describes the development of a distribution function over the set of charge occupation numbers. The driving force behind this evolution are charge tunneling events among nanoparticles and their associated rates. In this paper, we introduce two meanfield approximations to this master equation. By parametrization of the distribution function using its first- and second-order statistical moments, and a subsequent projection of the dynamics onto the resulting moment manifold, one can deterministically calculate expected charges and currents. Unlike a kinetic Monte Carlo approach, which extracts samples from the distribution function, this meanfield approach avoids any random elements. A comparison of results between the meanfield approximation and an already available kinetic Monte Carlo simulation demonstrates great accuracy. Our analysis also reveals that transitioning from a first-order to a second-order approximation significantly enhances the  accuracy. Furthermore, we demonstrate the applicability of our approach to time-dependent simulations, using eulerian time-integration schemes.
\end{abstract}

\maketitle


\section{Introduction}
Disordered networks of $\mathrm{Au}$-nanoparticles \cite{nnano}, interconnected by organic molecules, exploit physical processes for computation. Each nanoparticle (NP) is a conductive island and enables charge tunneling between neighboring islands when sufficient differences in free energy exist \cite{van2002electron, wasshuber2002single}. The network is surrounded by electrodes categorized as input, output, or control electrodes. Applying voltages to the control electrodes allows modifying the internal charge tunneling dynamics, resulting in varying input-output relationships. Consequently, one can interpret the set of applied control electrode voltages as the genome of the device. Tuning it through genetic algorithms or backpropagation \cite{Backprop}\cite{nnano}, one can seek configurations in which the system can accomplish a given task, while being fast, energy-efficient and reconfigurable.

The phase space of the system encompasses all possible charge occupation numbers within the network. Its charge tunneling dynamics are defined by the master equation, formulated as a set of differential rate equations \cite{fonseca1995numerical, willy2016modeling}. The time evolution of these equations can be solved with a Kinetic Monte Carlo (KMC) algorithm \cite{fonseca1995numerical, wasshuber2002single, mensing2023kinetic, 658562}. This approach efficiently reduces the infinite phase space to a physically meaningful subset, despite its intrinsic stochastic nature. Attempting to formulate an exact deterministic solution for the master equation, one faces exponential computational complexity arising from the size of phase space. Consequently, exact solutions are practically limited to very small systems. A strategy to formulate a deterministic solution with manageable polynomial complexity involves the application of mean-field approximation. In this context, a first-order mean-field approximation for charge tunneling dynamics was introduced in \cite{Lawrence}. These ideas have also been used to solve the master equation in socioeconomic population dynamics and migration processes \cite{haag2017modelling, haag1984stochastic, weidlich1986stochastic, haag1983toward}. Besides similar concepts have also been used to describe triplet-triplet annihilation in phosphorescent emission layers of light-emitting diodes \cite{PhysRev}.  

In this work, we present a systematic approach that extends previous mean field approximations to higher orders. We apply the mean field approximation to the master equation describing single electronic tunneling dynamics. However, these concepts should also be applicable to any kind of stochastic processes where the state evolves in time by migrating single discrete values. Specifically, we explore the second-order method, demonstrating its superior performance over the first-order method while maintaining high efficiency. An illustrative step towards analyzing time-dependency and the investigation of nonlinear input/output electrode relationships is undertaken.

\section{Theoretical Background}
\begin{figure}
    \centering
    \includegraphics[width=0.35\textwidth]{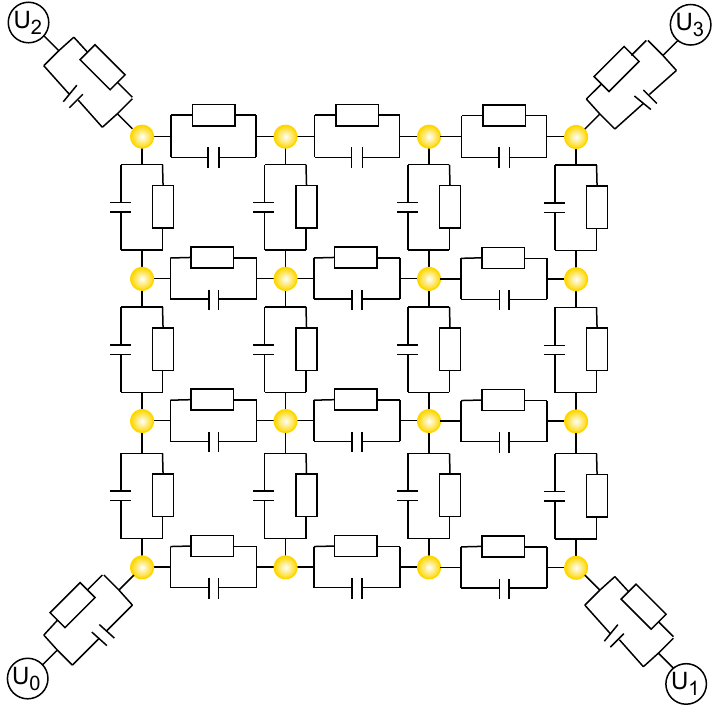}
    \caption{A $4\times4$ nanoparticle network with four electrodes $U_0$, $U_1$, $U_2$, and $U_3$. The electrodes are attached at the corners. The tunnel junctions are depicted as capacitors and resistors in parallel.}
    \label{fig:44_net_sketch}
\end{figure}

The network is modelled as a collection of $N$ gold-NPs arranged in a grid lattice, some of which are connected to electrodes (see FIG. \ref{fig:44_net_sketch}). The occupation number \mbox{$n \in \mathbb{Z}$} represents the number of excess electrons residing on a NP (island). This description will suffice to build the phase space $\Omega$. For $N$ particles, the phase space is thus

\begin{equation}
    \Omega = \mathbb{Z}^N.
\end{equation}

The state of the system is described by a state vector $\vec{n} \in \Omega$ which contains occupation numbers for excess electrons on each island. The network topology is encoded in a capacitance matrix, linking nearest neighbors by their mutual capacitance. The internal electrostatic energy can thus be calculated. Due to their proximity, electrons can tunnel through junctions between nearest neighbors. The tunnel rates will be described according to a zero-dimensional \emph{orthodox tunnel theory} \cite{AVERIN1991173}. The tunnel rate $\Gamma$ for an electron from one position to another depends on the difference in Helmholtz free energy $\Delta F$ associated with this tunnel process

\begin{equation}
    \Gamma = - \frac{\Delta F}{e^2 R} \cdot \left[ 1 - \mathrm{exp\,\frac{\Delta F}{k_B T}} \right]^{-1}.
\end{equation}

We restrict our model to single-electron-tunneling. For a more complete description of single-electron tunneling applied to nanoparticle networks, we refer to \cite{mensing2023kinetic}.

It is important to mention that the free energy $\Delta F$ governing the underlying transition rate not only depends on the population of the two involved islands but rather on the complete state vector $\vec{n}$ due to the network of capacitors.  

We introduce a distribution function $\rho$ over phase space $\Omega$ which
assigns a probability $\rho(\vec{n})$ to each possible system state $\vec{n}$ \cite{MasterEq, Lawrence}. The tunnel rates describe the tunneling of electrons between islands or electrodes and can be interpreted as the transition rate from states $\vec{n}$ to $\vec{m}$, denoted as $\Gamma_{\vec{n}\,\vec{m}}$. Hence, the rate for an electron to tunnel from island $i$ to island $j$ is the transition rate from state $\vec{n}$ to state $\vec{n} - \vec{e_i} + \vec{e_j}$, where the latter is missing one electron at index $i$ and having an additional one at index $j$. $\vec{e_i}$ is the i-th basis vector in $\Omega$. If an electrode is connected to island $i$, two more transition rates are possible: For an electron tunneling towards the island, a transition from state $\vec{n}$ to state $\vec{n} + \vec{e_i}$, the rate $\Gamma_{\vec{n}\,\vec{n} + \vec{e_i}}$ is associated. For the reverse process, a minus sign must be taken. All other transition rates are zero.

The dynamics of the distribution function are governed by the master equation:
\begin{equation}
    \label{eq:master_eq}
    \partial_t\, \rho(\vec{n}) = \sum_{\vec{m} \neq \vec{n}}( \Gamma_{\vec{m}\,\vec{n}} \,\rho (\vec{m}) - \Gamma_{\vec{n}\,\vec{m}} \,\rho (\vec{n})).
\end{equation}
The equilibrium distribution function is found when \mbox{$\partial_t \, \rho(\vec{n}) = 0 \;\forall \vec{n}$}. Afterward, expectation values of quantities of interest can be taken. From a nanoparticle the electron can hop to another nanoparticle or to an attached electrode (or vice versa). This is captured by the following phase-space functions: The current flowing from particle $i$ to particle $j$
\begin{align}
    \label{eq:inter_particle_current}
    I_{ij}(\vec{n}) &= \Gamma_{\vec{n}\,\vec{n} + \vec{e_j} - \vec{e_i}} - \Gamma_{\vec{n}\,\vec{n} - \vec{e_j} + \vec{e_i}}
\end{align}
the current flowing towards particle $i$ from its attached electrode
\begin{align}
    \label{eq:electrode_current}
      I_{ei}(\vec{n}) &= \Gamma_{\vec{n}\,\vec{n}+\vec{e_i}} - \Gamma_{\vec{n}\,\vec{n}-\vec{e_i}}
\end{align}
and the total current flowing to particle $i$
\begin{align}
    \label{eq:total_current}
    I_i(\vec{n}) = \sum_{j \neq i}I_{ji}(\vec{n}) + I_{ei}(\vec{n})
\end{align}
Furthermore, phase-space functions with opposite sign are introduced, which are useful later
\begin{align}
     I_{ij}^\dag(\vec{n}) &= \Gamma_{\vec{n}\,\vec{n} + \vec{e_j} - \vec{e_i}} + \Gamma_{\vec{n}\,\vec{n} - \vec{e_j} + \vec{e_i}} \\
     I_{ei}^\dag(\vec{n}) &= \Gamma_{\vec{n}\,\vec{n}+\vec{e_i}} + \Gamma_{\vec{n}\,\vec{n}-\vec{e_i}} \\
     I_i^\dag(\vec{n}) &= \sum_{j \neq i}I_{ji}^\dag(\vec{n}) + I_{ei}^\dag(\vec{n})
          \label{eq:I_dagger}
\end{align}
For the subsequent equations, we use the convention that $I_{ei}=0$ if there is no direct connection between island $i$ and an electrode.

Since storing and updating the entire distribution function is intractable for large systems, we will introduce efficient mean field algorithms.  For this purpose, equations of motion for the first and second moment are derived. With the definition of the first moment 
\begin{align}
    \langle n_i \rangle = \sum_{\vec{n}}n_i\, \rho(\vec{n})
\end{align}
its time evolution can be expressed as
\begin{align}
    \partial_t \langle n_i \rangle = \sum_{\vec{n}}n_i\, \partial_t \rho(\vec{n}).
\end{align}
Inserting equation (\ref{eq:master_eq}) leads to
\begin{align}
    \partial_t \langle n_i \rangle = \sum_{\vec{n}}n_i\,\sum_{\vec{m} \neq \vec{n}}( \Gamma_{\vec{m}\,\vec{n}} \,\rho (\vec{m}) - \Gamma_{\vec{n}\,\vec{m}} \,\rho (\vec{n}))
\end{align}
Expanding the sum to express the equation as an expectation value, one obtains
\begin{align}
\label{eq:expectation_head}
    \partial_t \langle n_i \rangle = \sum_{\vec{v}} \rho(\vec{v})\sum_{\vec{n}\neq\vec{v}} (n_i - v_i) \Gamma_{\vec{v}\,\vec{n}}.
\end{align}
Since the majority of transition rates $\Gamma$ is zero because of \emph{single electron tunneling}, this expression simplifies to a very descriptive form. For a given $\vec{v}$, $\Gamma_{\vec{v}\,\vec{n}}$ is non-zero if \mbox{$\exists\, i, j $} with $ i \neq j$ and \mbox{$\vec{n} = \vec{v} \pm \vec{e_i} \mp \vec{e_j}$} for inter-particle tunneling. If an electrode is attached to island $i$, two further transition rates, caused by the electrodes, can be non-zero: $\Gamma_{\vec{v}\,\vec{v}\pm\vec{e_i}} \neq 0$. Furthermore, all rates where $n_i$ remains unchanged are not accounted for in the summation above, since $(n_i - v_i) = 0$ in that case. Equation (\ref{eq:expectation_head}) simplifies to
\begin{align}
    \partial_t \langle n_i \rangle
    &=  \sum_{j \neq i} \langle I_{ji} \rangle + \langle I_{ei} \rangle \\
    \label{eq:first_order_equation}
    &= \langle I_i \rangle
\end{align}
using equation (\ref{eq:electrode_current}) and (\ref{eq:inter_particle_current}). This equation just reflects the conservation of electrons.

Analogously, the dynamics of higher order moments are derived. For the second order moments $\langle n_i^2 \rangle$ one can generalize equation (\ref{eq:expectation_head}) to
\begin{align}
    \partial_t \langle n_i^2 \rangle = \sum_{\vec{v}} \rho(\vec{v})\sum_{\vec{n}\neq\vec{v}} (n_i^2 - v_i^2) \Gamma_{\vec{v}\,\vec{n}}
\end{align}
Again, when taking care about the possible transitions, in analogy to the first order method, one ends up after a straightforward calculation with
\begin{equation}
\begin{split}
  \partial_t \langle n_i^2 \rangle = \sum_{j \neq i} \left( 2\langle n_i I_{ji} \rangle + \langle I_{ji}^\dag \rangle \right) \\+\, 2\langle n_i I_{ei} \rangle + \langle I_{ei}^\dag \rangle 
\end{split}
\end{equation}
This can be abbreviated as
\begin{equation}
    \partial_t \langle n_i^2 \rangle =  \langle 2 n_i I_{i} \rangle + \langle  I^\dag_{i} \rangle
    \label{eq:second_order_equation}
\end{equation}
This procedure can be easily generalized to either mixed correlation terms such as $\langle n_i n_j \rangle $ or higher order moments. Due to the combination of simplicity and accuracy, observed after incorporation of the $\langle n_i^2 \rangle $ correlation, in this work we restrict ourselves to the simplest extension beyond the first-order case. 

Taking the term $ I_{ji}^\dag$ is an example one might be tempted to conclude that this term only depends on the populations of islands $i$ and $j$. However, this is not true since the free energy $\Delta F$, governing the underlying transition rates, depends on the populations of all other sites

Note, that all the equations so far are exact but intractable since the expectation values still require knowledge of the full exponentially complex distribution function. The mean field approximation will be introduced in the next section, which allows one to express the expectation values just in terms of the first and (possibly) second moment.  

\section{Numerical Methods}
Strictly speaking, calculation of, e.g., $\langle I_{ij} \rangle$ requires knowledge of the full distribution $\rho(\vec{n}, t)$ which requires the solution of the exact master equation (which we want to avoid). Let us assume that we need to evaluate an expectation value which explicitly depends on islands $i$ and $j$ such as  $\langle I_{ij} \rangle$. Then we can proceed as follows:

\textit{a. Mean-field approximation}
Following \cite{Lawrence} we neglect all fluctuations of the electron occupation numbers $n_k$ where the index $k$ reflects all indices which fulfill $k\ne i$ and $k \ne j$. The $\{n_k\}$ are substituted by their average values $\{\langle n_k \rangle \}$. This is a natural choice because the transition rates from cluster $i$ to cluster $j$ will naturally most strongly depend on the number of electrons on both clusters whereas one may hope that the remaining clusters only act via their average electron occupation. Then the expectation value $\langle I_{ij} \rangle$ just requires an average over the two-dimensional distribution   $\widetilde{\rho}(n_i,n_j)$. This average is now abbreviated as $\langle I_{ij} \rangle_{i,j}$.\par

\textit{b. Factorization}
Furthermore, we neglect correlations among adjacent clusters. Preliminary KMC simulations have shown that these correlations are small. This step strongly simplifies the realization of the next step. Thus, we write $\widetilde{\rho}(n_i,n_j) = \widetilde{\rho}_i(n_i) \cdot \widetilde{\rho}_j(n_j)$. To simplify the notation, we just abbreviate each factor as $\widetilde{\rho}(n) $ with $n \in \{i,j\}$. When an island is directly connected to an electrode, also terms with just one index occur. For these terms, no factorization is required.\par

\textit{c1. Choice of $\widetilde{\rho}(n)$ based on knowledge of the first moment (meanfield-1 {\it MF1})}

If only information about the first moment is available, a natural choice, introduced in \cite{Lawrence}, reads
\begin{align}
\label{eq:lawrence_dist_1d}
    \widetilde{\rho}(n) =
                    \begin{cases}
                        d &\text{for}\quad n = \lceil \langle n \rangle \rceil \\
                        1 - d  &\text{for}\quad n = \lfloor \langle n \rangle \rfloor \\
                        0 &\text{otherwise}
                    \end{cases}
\end{align}
with $d = \langle n \rangle -  \lfloor \langle n \rangle \rfloor$. Notably, this distribution has a mean $\langle n \rangle$ and only assigns probabilities to the two occupation numbers adjacent to its mean. Using those approximations, equation (\ref{eq:first_order_equation}) becomes closed for $N$ particles. The variance reads $d - d^2$. This is the smallest possible variance of the electron occupation number.\par

\textit{c2. Numerical determination of $\widetilde{\rho}(n) $ based on additional information about the second moment (meanfield-2 {\it MF2})} 
Incorporating also information about the second moment allows for a more realistic approximation of $\widetilde{\rho}(n)$. Here, two different approaches are presented. First, we define $\widetilde{\rho}(n)$ for all $n \in [-20,20]$ and guarantee that  $\widetilde{\rho}(n)$ has the correct first and second moments. However, the choice is not unique. Therefore, we additionally formulate an optimization problem that the distribution 
 $\widetilde{\rho}$ shall maximize the entropy $S = -\sum_{n}\widetilde{\rho}(n) \log \widetilde{\rho}(n)$ under boundary conditions:
\begin{equation}
\begin{aligned}
\label{eq:boundary_cond_gauss_1d}
    G_0 &= \sum_{n}\widetilde{\rho}(n) - 1 &= 0 \\
    G_1 &= \sum_{n}n\widetilde{\rho}(n) - \langle n \rangle &= 0 \\
    G_2 &= \sum_{n}n^2\widetilde{\rho}(n) - \langle n^2 \rangle &= 0 
\end{aligned}
\end{equation}
The solution can be directly obtained with the use of Lagrange parameters, yielding the discrete Gaussian distribution
\begin{align}
    \widetilde{\rho}(n) = \frac{1}{Z}e^{-\frac{(n - \mu)^2}{2\sigma^2}} \quad\text{for}\quad n \in \mathbb{Z}.
\end{align}
The Lagrange multipliers $Z$, $\mu$ and $\sigma$ are determined numerically so that all three boundary conditions (\ref{eq:boundary_cond_gauss_1d}) are fulfilled.\par

\textit{c3. Analytical determination of $\widetilde{\rho}(n) $ based on additional information about the second moment (quick meanfield-2 {\it QMF2})}
For two reasons the implementation MF2 is relatively slow. First, the distribution function has to be evaluated for a large phase space, second, a numerical determination of the Lagrange parameters is required. Here we suggest a method which remedies both problems. First, we restrict ourselves to a phase space size of 4, i.e. we have to determine four probabilities $p_1,...,p_4$. Here the four indices reflect subsequent natural numbers such that the two lower ones ($n_1,n_2$) are below or equal to the current estimate $\langle n \rangle $ and the other two ($n_3,n_4$) just above. Second, rather than maximizing the standard Shannon entropy we maximize
\begin{equation}
    \tilde{S} = 1 - \sum_{m=1}^4 p_m^2 . 
\end{equation}
Together with the boundary conditions\begin{align}
    \label{eq:p24_boundary_1}
    \sum_{m=1}^{4} p_m &= 1 \\
    \label{eq:p24_boundary_2}
    \sum_{m=1}^{4} n_m  p_m &= \langle n \rangle \\
    \label{eq:p24_boundary_3}
    \sum_{m=1}^{4} n_m^2 p_m &= \langle n^2 \rangle
\end{align}
we obtain after a straightforward calculation
\begin{equation}
    \spalignsysdelims{.}{.}
    \spalignsys{
     p_1 = \frac{1}{4}d^2 - \frac{11}{20}d + \frac{1}{4}(\Delta n)^2 + \frac{3}{20} ;
     p_2 = -\frac{1}{4}d^2 + \frac{3}{20}d - \frac{1}{4}(\Delta n)^2 + \frac{11}{20} ;
     p_3 = -\frac{1}{4}d^2 + \frac{7}{20}d - \frac{1}{4}(\Delta n)^2 + \frac{9}{20} ;
     p_4 = \frac{1}{4}d^2 + \frac{1}{20}d + \frac{1}{4}(\Delta n)^2 - \frac{3}{20}
     }
\end{equation}
Thereby, $d = \langle n \rangle - \lfloor \langle n  \rangle \rfloor$ and $(\Delta n)^2$ is the current estimate of the variance. The distribution will be referred to as $p^2$-4 distribution.

This specific choice of $ \tilde{S}$ can be rationalized in two ways. First, if $1 - p_n$ is not too large, $ \tilde{S}$ emerges from a Taylor expansion of the Shannon entropy. Second, in analogy to the Shannon entropy, $ \tilde{S}$ has the property that the maximum under the normalization condition $\sum_m p_m = 1$  reads $p_m = const$. Indeed, we later show that the resulting distribution $\widetilde{\rho}(n) $ hardly differs from using the Shannon entropy.

One technical issue needs to be discussed in more detail. Formally, the $p_i$ can turn out to be negative. This would invalidate the interpretation as probabilities. Here, we restrict this discussion to $d \le 1/2$. The results for $d > 1/2$ follow from symmetry arguments and the relevant results for $d > 1/2$ are found in the appendix \ref{appendix:p24_dist}.

From the general solutions one would obtain $p_3 < 0 $ if $(\Delta n)^2 > V_+ \equiv \frac{9}{5} + \frac{7}{5}d - d^2$ and $p_4 < 0$ if $(\Delta n)^2 < V_- \equiv \frac{3}{5} - \frac{1}{5}d - d^2$.
One can easily convince oneself that upon increasing or decreasing $(\Delta n)^2$ first $p_3$ or $p_4$ become negative, respectively, whereas $p_1$ and $p_2$ are still positive. 

Now a simple strategy can be formulated. In case of $(\Delta n)^2 > V_+$ one chooses $p_3 = 0$. The remaining $p_i$ are completely determined by the three boundary conditions. One obtains 
\begin{align}
    p_1 &= \frac{1}{3}(d^2 - 2d + (\Delta n)^2) \\
    p_2 &= \frac{1}{2}(-d^2 + d - (\Delta n)^2 + 2) \\
    p_4 &= \frac{1}{6}(d^2 + d + (\Delta n)^2)
\end{align}
In the regime $V_+ \le (\Delta n)^2 < 2 +d - d^2$ all $p_m$ are positive. In practical terms this means that the reduction to a four-dimensional phase space is consistent as long as the variance is smaller than 2 (in the worst case $d=0$). As shown for the specific examples below this is always fulfilled. Otherwise, it would be straightforward, to generalize these ideas to, e.g., a six-dimensional phase space.

In the other limit $(\Delta n)^2 < V_-$ one should choose $p_4 = 0$. Again, from consideration of the three boundary conditions one obtains
\begin{align}
    p_1 &= \frac{1}{2}(d^2 - d + (\Delta n)^2) \\
    p_2 &= -d^2 - (\Delta n)^2 + 1 \\
    p_3 &= \frac{1}{2}(d^2 + d + (\Delta n)^2)
\end{align}
All $p_i$ are non-negative as long as $(\Delta n)^2 \ge d - d^2$. 
Since $d - d^2$ is the minimum variance which may occur (see also the discussion of the MF1 algorithm) one can always find a well-defined distribution in the limit of low variances.

\begin{figure}
    \centering
    \includegraphics[width = 0.49\textwidth]{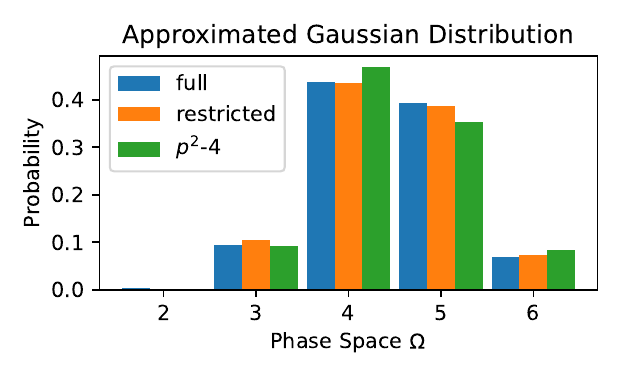}
    \caption{Different approximations to the full Gaussian distribution with $\langle n \rangle = 4.43$ and $(\Delta n )^2$ = 0.6 are shown. Beyond the distributions used in MF2 and QMF2, we have also included the distribution resulting from maximization of the Shannon entropy, also restricted to a four-dimensional phase space, denoted restricted Gaussian distribution. }
    \label{fig:restricted_gaussian}
\end{figure}

This procedure results in similar probabilities as the discrete Gaussian at minimal computational cost. A comparison is shown in FIG. \ref{fig:restricted_gaussian}. Indeed, using the $p^2$-4 distribution shows little deviations from the complete as well as from the restricted Gaussian distribution.

In summary, the dynamics are projected onto the moment manifold, which has a much lower dimensionality. The resulting dynamics of moments together with the respective mean field distribution serve as an approximation to the true dynamics.

We mention in passing that the time evolution of the correlation of two adjacent sites involves terms such as $\langle n_i I_j \rangle$. From the definition of $I_j$ it follows that three clusters are involved at the same time. Whereas for the QMF2 algorithm the tunneling rates have to be evaluated for 16 different cases $(n_i,n_j)$, an additional factor 4 would emerge when taking into account the cross-correlations.  

To solve for the equilibrium configuration $\partial_t \langle n_i \rangle = \partial_t \langle n_i^2 \rangle = 0$, one can integrate the differential equations and obtain the full time-resolved dynamics. Since this is usually not required, more efficient algorithms finding the equilibrium point can be used. The ADAM-algorithm \cite{ADAM}, commonly used in machine-learning to efficiently follow gradients of a cost function, can be applied and provides faster convergence and runtime performance by damping oscillations.

\section{Simulation Results}
Simulation results of first- and second-order methods are compared for systems of one or multiple nanoparticles. For one-particle systems, the exact distribution function can be obtained by solving the master equation as reference. The abilities of MF1, MF2 and QMF2 will be assessed and compared. Since solving the master equation becomes computationally too complex for larger systems, a kinetic Monte-Carlo method \cite{mensing2023kinetic} (KMC) is used for generating samples of the distribution function, and thus reference data for larger systems.  This method generates samples $\vec{n}^{(j)} \in \Omega$ for $j = 1, \ldots, M$ of the distribution function $\rho$. The first- and second-order moments of the distribution can then be obtained by an estimate:
\begin{align}
    \langle n_i \rangle &= \frac{1}{M}\sum_{j = 1}^M n_i^{(j)} \\
    \langle n_i^2 \rangle &= \frac{1}{M}\sum_{j = 1}^M \left(n_i^{(j)}\right)^2
\end{align}
The currents involving electrodes are calculated by counting the number of jumps $\vec{n} \rightarrow \vec{n}\pm\vec{e_i}$ for a given amount of time.

\subsection{Single Electron Transistor}
\begin{figure}
    \centering
    \includegraphics[width = 0.35\textwidth]{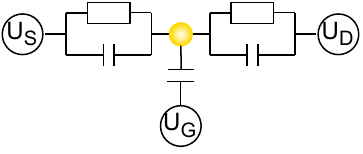}
    \caption{A circuit diagram of the single-electron-transistor. To the left and right, the two connected electrodes (source and drain) and the gate, realized by the capacitively coupled silicon substrate, are shown. Tunnel junctions are indicated as capacitors and resistors in parallel.}
    \label{fig:set_circuit}
\end{figure}

The single electron transistor is an important example of a network architecture consisting just of one nanoparticle. It is placed on a silicon substrate and connected to two electrodes who act as an input and output. Furthermore, a gate voltage can be applied to the silicon substrate, allowing for further manipulation of the potential landscape. A circuit diagram is shown in FIG. \ref{fig:set_circuit}.

\begin{figure}
    \centering
    \includegraphics[width = 0.49\textwidth]{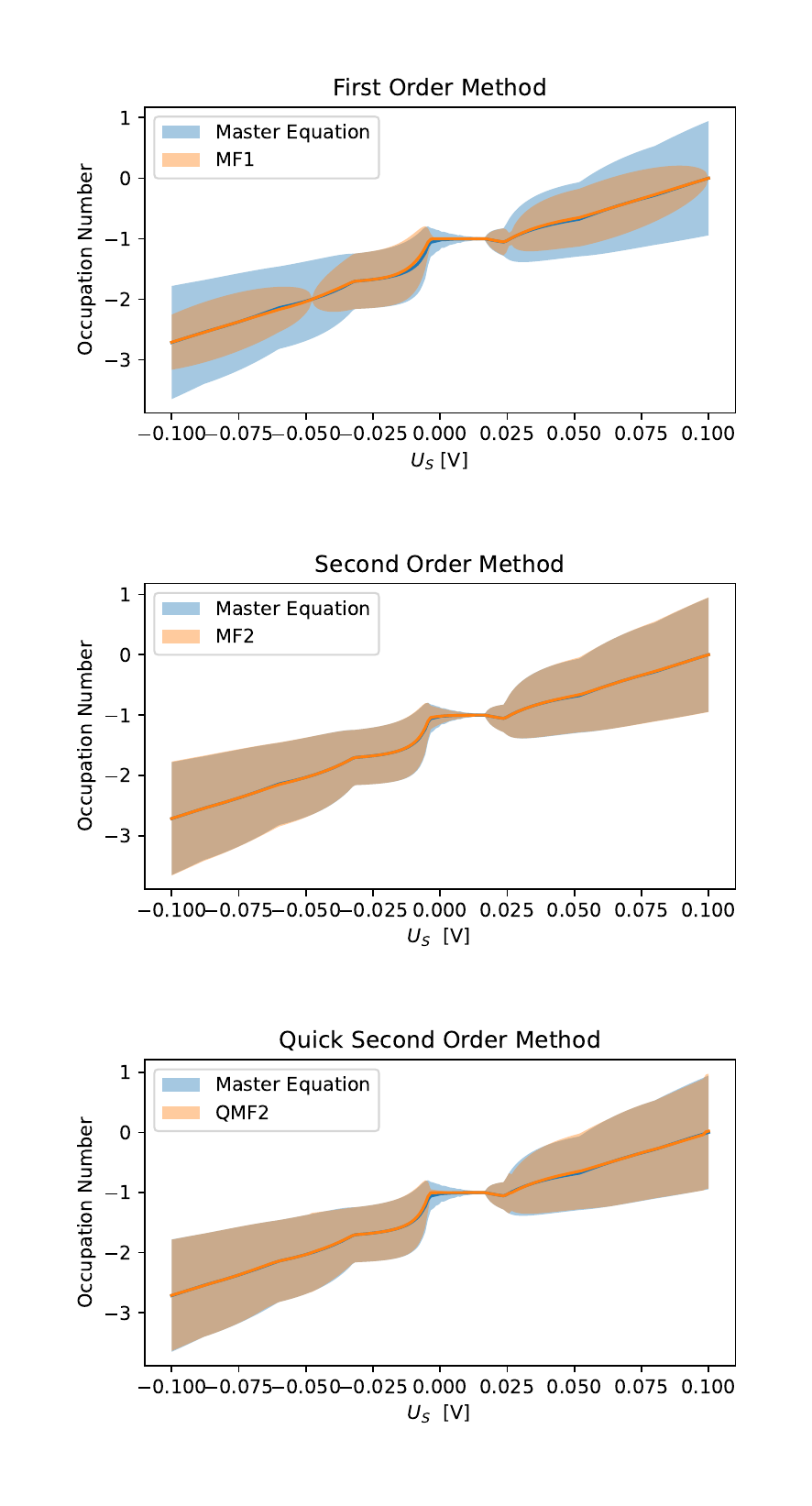}
    \caption{For a non-zero gate voltage of $0.05\,\si{V}$, the first order method has problems replicating the mean occupation number and its variance. The second order methods still accurately reproduce the mean and variance of the master equation. The shaded areas represent the interval of $\pm\Delta n$ predicted by the methods compared to the master equation. It can be seen that for low input voltages, no charge resides on the island, a non-linear effect called Coulomb blockade (see \cite{mensing2023kinetic}).}
    \label{fig:set_3_methods}
\end{figure}

The calculated mean occupation number $\langle n \rangle$ and its predicted standard deviation $\Delta n$ are displayed in comparison to the correct master equation solution in FIG. \ref{fig:set_3_methods}.
The first order distribution (equation (\ref{eq:lawrence_dist_1d})) cannot account for high variances, which naturally appear in the master equation. The second order method MF2, using the full Gaussian distribution, does surpass this problem and accurately reproduces mean and variance of the occupation number. The QMF2 methods using the $p^2$-4 distribution performs similarly to the MF2 method, thus providing a much closer approximate to the true distribution function and the ability to calculate more exact expectation values (like the output current) or even evaluate the standard deviation of those.

\subsection{Multi-Particle Systems}
As a more relevant example we consider square-shaped networks of nanoparticles of different sizes $l\times l\times1$ for $2 \leq l \leq 10$, exploring system sizes of up to a hundred nanoparticles, which enter the range of experimentally realized systems. Four electrodes are attached to the corners of the network, providing a variety of input voltages, as well as a gate voltage. The systems are equilibrated and currents at electrode 3, being signified the output-electrode, are calculated. An example of such a system is shown in FIG. \ref{fig:44_net_sketch}. While the system size is varied from $2 \times 2 \times 1$ up to $10 \times 10 \times 1$, the electrode positions stay at the corners. As the MF2 method is numerically quite expensive for large systems, the focus lies on the QMF2 method.

\begin{figure}
    \centering
    \includegraphics[width = 0.4\textwidth]{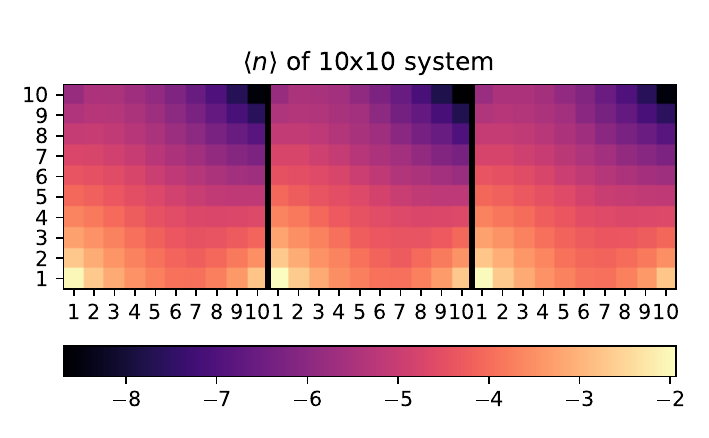}
    \caption{For a random voltage configuration, the mean occupation numbers are plotted from left to right for the KMC, first order MF1 and second order QFM2 method. A continuous gradient can be seen. Results of all methods lie in qualitative agreement. $E_1 = -0.10\,\si{V}$, $E_2 = -0.23\,\si{V}$, $E_3 = 0.07\,\si{V}$, $E_4 = 0.05\,\si{V}$ are set and the gate voltage is $G = 0.13\,\si{V}$.}
    \label{fig:visual_states10x10}
\end{figure}

Obtaining the equilibrium occupation numbers for the KMC-, MF1-, and QMF2-algorithm and plotting them in their spatial configuration (here for a 10 by 10 system) results in FIG. \ref{fig:visual_states10x10}.

\subsubsection{Accuracy Comparison}
The main improvement of the second order method lies in the estimation of electrode currents, which are the main point of interest when mapping functionalities to input output electrode dependencies. Compared to the KMC reference-data, whose currents are sampled until a relative error of $0.5\%$, requiring an average amount of $74\cdot 10^6$ samples, the first order method estimates the currents with a relative error of $(10.47 \pm 0.5)$\% on average over all system sizes, which is outperformed by the second order method with an error of $(3.25 \pm 0.5)$\%, shown in table \ref{tab:relative_errors_currents}. As the estimation of mean occupation numbers behave similarly in MF1 and QMF2, the improved output current estimation arises through the additional incorporation of the second moment.

\begin{table}
\label{tab:relative_errors_currents}
 \caption{Relative Errors in Output Current [$\pm 0.5$\%]}
\begin{ruledtabular}
\begin{tabular}{ccc}
        System Size  & MF1 [\%] & QMF2 [\%]\\
        4 & 6.72 & 2.89 \\
        9 & 7.81 & 2.29 \\
        16 & 9.20 & 2.50 \\
        25 & 10.53 & 3.02 \\
        36 & 11.66 & 3.44 \\
        49 & 12.04 & 3.29 \\
        64 & 12.39 & 4.14 \\
        81 & 12.40 & 4.62 \\
        100 & 11.50 & 5.52 \\
\end{tabular}
\end{ruledtabular}
\end{table}

\subsubsection{Predicted Means and Variances}
\begin{figure}
    \centering
    \includegraphics[width = 0.49\textwidth]{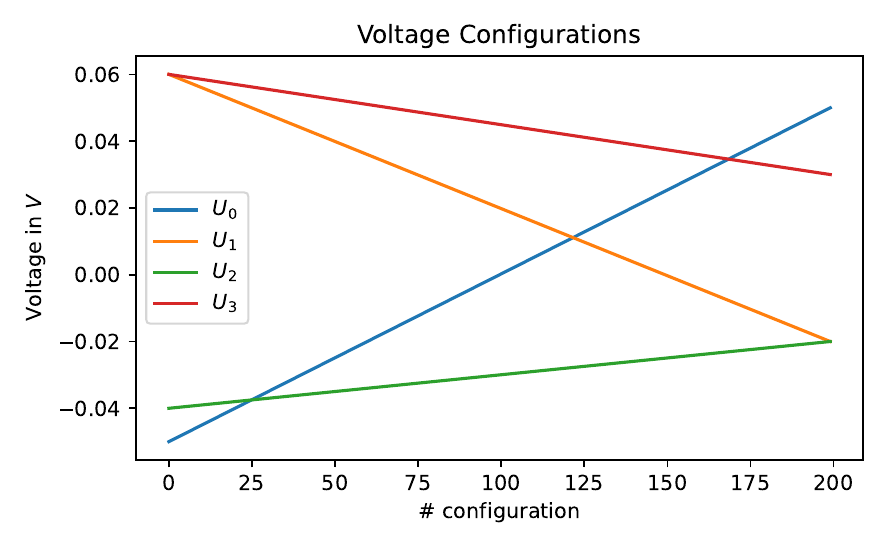}
    \caption{200 voltage configurations for the 4 electrodes ($U_0$ to $U_3$) are taken continuously to show the response of the system.}
    \label{fig:contiuous_voltages}
\end{figure}

\begin{figure}
    \centering
    \includegraphics[width = 0.49\textwidth]{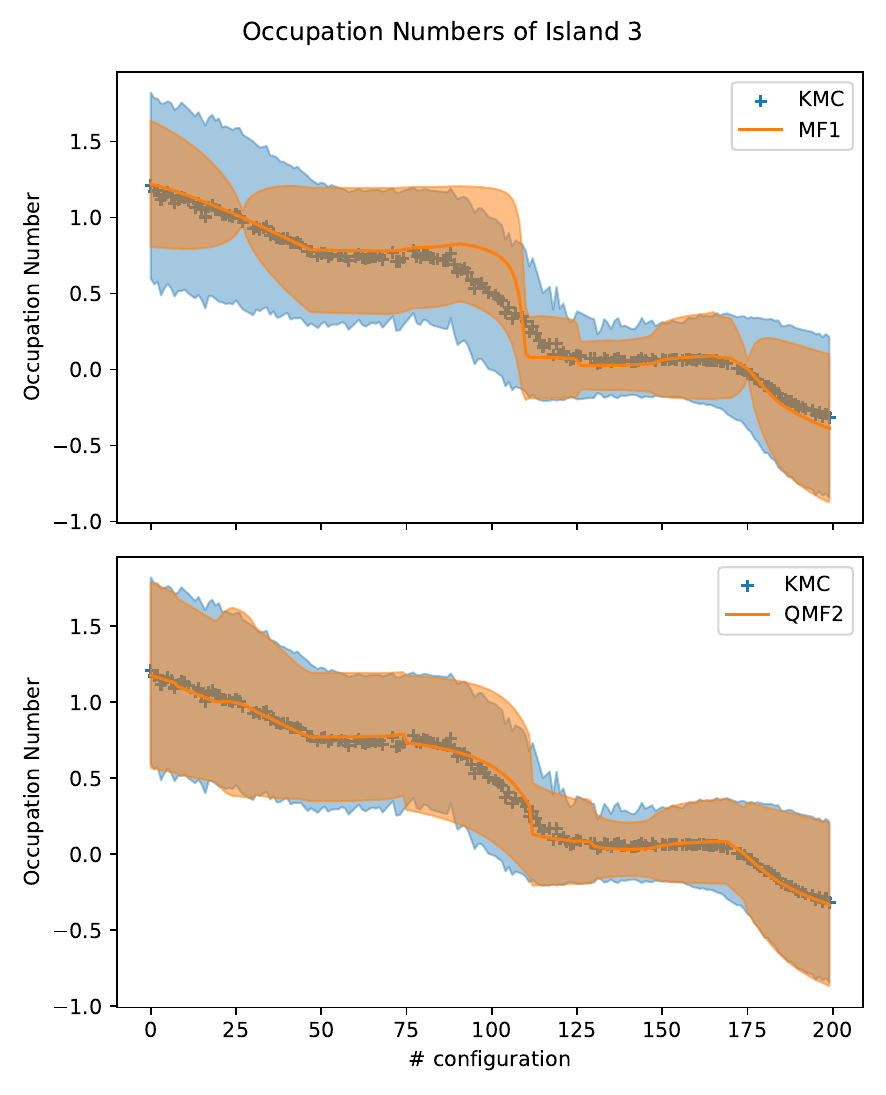}
    \caption{The mean and the interval of $\pm \Delta n$ resulting from first- and second-order method are shown for the island connected to electrode $U_0$. The KMC means are highlighted as red crosses, and the red shaded area corresponds to their standard deviation. It can be seen that the second order method achieves a better approximation than the first order method, thus allowing for more precise estimation of expectation values.}
    \label{fig:cont_comparison_island0}
\end{figure}

To evaluate the response to adiabatically changed voltage configurations, 200 continuous voltages, shown in FIG. \ref{fig:contiuous_voltages}, are applied to a 4 by 4 system, as used above. The predicted quantities $\langle n_i \rangle$, $\langle n_i ^2 \rangle$ and their deviations in different algorithms are depicted as the system is equilibrated for each voltage configuration.
See FIG. \ref{fig:cont_comparison_island0} for the island connected to electrode $U_0$. Naturally, the second order method QMF2 more closely matches the variances. However, importantly for some configurations (e.g. around 130-140) also the first moment is predicted much better although for this specific island in this regime the variance is close to is minimum value, i.e. seeing similar results for MF1 and QMF2. This exemplifies the superiority of QMF2, even if one is only interested in the first moments. 

\subsection{Time Dependent Systems}
\begin{figure}
    \centering
    \includegraphics[width = 0.5\textwidth]{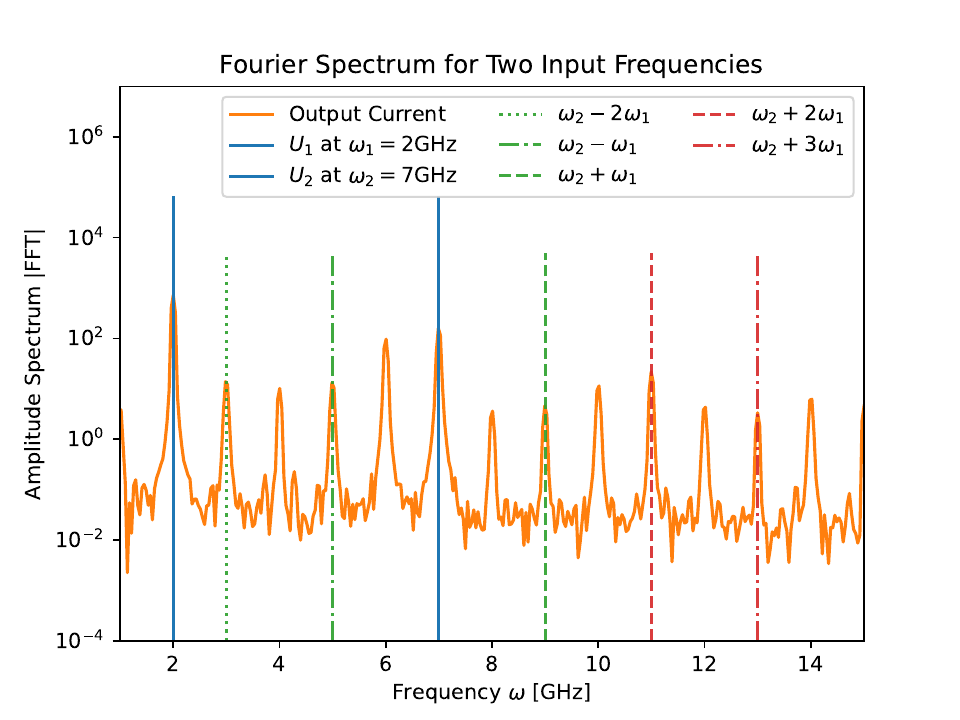}
    \caption{Subjecting the system to two different input signals at different electrodes (diagonals) the output current not only includes their frequencies but also a nonlinear response consisting of linear combinations of those frequencies.}
    \label{fig:time_dep_multiplic}
\end{figure}

The algorithms at hand not only allow for equilibrium-state analysis, but can be easily used to calculate the system's response to time-dependent voltage configurations. A Fourier analysis is conducted to quantify different types of non-linearities characteristic to the system. The system in question is a 3 by 3 system (analogous to the systems above) with two input electrodes attached at the diagonals. Output electrodes are attached to the remaining corners with a voltage set to zero. Two cosine-voltages are applied with amplitude $U_0 = 0.1\,\si{V}$ and different frequencies $\omega_1 = 2\,\si{GHz}$ and $\omega_2 = 7\,\si{GHz}$.
A nonlinear frequency mixture is observed when the system is subjected to the two different oscillating input voltages. The inputs couple and produce linear combinations of their frequencies in the output spectrum, showing that the system is in practice capable of multiplying two signals at different electrodes. See FIG. \ref{fig:time_dep_multiplic}. This behavior corresponds to an AND logic-gate. Further tuning of control voltages could result in more complex nonlinear behavior (e.g. calculating exponentials, acting as boolean logic-gates or performing high-dimensional input-output transformations).

\section{Conclusion}
For enabling an appropriate analysis of the system's properties to ultimately exploit the inherent physical phenomena for computation, the system's dynamics have to be simulated. A corresponding master equation governs the dynamics as a probability-flow in the discrete phase space. This is infeasible to solve directly because the dimensionality of the phase space grows exponentially with the number of particles. Considering just two possible occupation numbers 0 and 1 for 100 islands would already exceed the capabilities as $2^{100}$ configurations result. Beyond the standard KMC-approach, we aimed to find methods to explore the dynamics deterministically. Conceptually, this in analogous to expressing the solution of a Langevin equation via a Fokker-Planck equation. For this purpose, guided by \cite{Lawrence}, we have derived equations governing the dynamics of the distribution's moments. Finding appropriate mean field distributions by maximization of entropy and substituting the true distribution led to a class of mean field algorithms. Its first- and second-order methods were presented. The incorporation of higher order moments or information about covariances could further improve the accuracy, albeit with the disadvantage of longer simulation times. Already when transitioning from first- to second-order, a great improvement of accuracy is observed, enabling sufficient analysis of the system's properties.

We remind that the Coulomb blockade of each nanoparticle is essential as it makes the system a concatenation of nonlinear switches which create its ability to map from input to outputs non-linearly with rich and reconfigurable dynamics. These effects can easily be reproduced with the presented algorithms. Furthermore, input/output relationships can be computed, be it adiabatic or time dependent. Time-dependent investigations, which by simple Eulerian integration are easily obtained, were performed to show that the system non-linearly couples the inputs of different electrodes, which is a prerequisite of evolving versatile functionalities.

Further investigations are needed for understanding the effects of covariance between occupation numbers, which in the aforementioned algorithms was neglected. A further increase in accuracy could be expected at the cost of a longer runtime. The covariances could be key to understanding the systems' behavior for very low input voltages, for which all current algorithms have difficulty. Also, a C++ accelerated implementation would allow for quicker and more effective research. Using frameworks like \emph{TensorFlow} or \emph{JAX}, an auto-differentiable implementation of the algorithm could be created to allow for identification of appropriate voltage configurations for difficult tasks, as backpropagation could be used and genetic algorithms become inefficient in high dimensional spaces. It could be attempted to evolve the system into classifiers or perform regression problems, as it is commonly done to benchmark algorithms and network architectures in Deep Learning. The flexibility of the system to change its input/output-relationship could thus be assessed.  

\begin{acknowledgments}
This work was funded by the Deutsche Forschungsgemeinschaft (DFG, German Research Foundation) through project 433682494--SFB 1459. We would like to thank P. Bobbert and W.G. van der Wiel for helpful discussions.
\end{acknowledgments}


\bibliography{apssamp}

\appendix

\section{$p^2$-4 Distribution}
\label{appendix:p24_dist}
 For $p_1 = 0$ the other probabilities amount to:
\begin{align}
 p_2 &= \frac{1}{2}d^2 -\frac{3}{2}d + \frac{1}{2}(\Delta n)^2 +1 \\
 p_3 &= -d^2 + 2d - (\Delta n)^2 \\
 p_4 &= \frac{1}{2}d^2 - \frac{1}{2}d + \frac{1}{2}(\Delta n)^2
\end{align}
Setting $p_2 = 0$, the other probabilities amount to:
\begin{align}
    p_1 &= \frac{1}{6}(d^2 - 3d + (\Delta n)^2 +2) \\
    p_3 &= \frac{1}{2}(-d^2 + d - (\Delta n)^2 + 2) \\
    p_4 &= \frac{1}{3}(d^2 + (\Delta n)^2 -1)
\end{align}

\end{document}